\documentclass[10pt,letterpaper]{article}
\usepackage[top=0.85in,left=1.00in,footskip=0.75in]{geometry}
\usepackage[utf8]{inputenc}
\usepackage{amsmath,amssymb}
\usepackage{algorithm,setspace}
\usepackage{algpseudocode}

\usepackage{mathtools}

\usepackage{changepage}

\usepackage{color}
\usepackage[table,xcdraw,dvipsnames]{xcolor}
\definecolor{cuteBlue}{rgb}{0.258, 0.387, 0.574}
\definecolor{color1}{rgb}{0.0, 0.16, 0.67}
\definecolor{color2}{rgb}{0.3, 0.0, 0.3}
\definecolor{color3}{rgb}{0.0, 0.33, 0.0}

\newcommand{\abs}[1]{\lvert#1\rvert}
\newcommand{\norm}[1]{\lVert#1\rVert}

\usepackage{nameref}
\usepackage[colorlinks=true, urlcolor=cuteBlue, citecolor=black, linkcolor=black]{hyperref}

\usepackage{microtype}
\DisableLigatures[f]{encoding = *, family = *}

\usepackage[aboveskip=5pt,font=small,labelfont=bf,labelsep=period,justification=raggedright,singlelinecheck=off]{caption}

\usepackage{xparse}

\DeclareDocumentCommand \eref{oooo} {\IfNoValueTF{#2}{Eq.~\ref{#1}}{\IfNoValueTF{#3}{Eqs.~\ref{#1} and \ref{#2}}{\IfNoValueTF{#4}{Eqs.~\ref{#1}-\ref{#3}}{Eqs.~\ref{#1}-\ref{#4}}}}}

\DeclareDocumentCommand \tref{oooo} {\IfNoValueTF{#2}{Table~\ref{#1}}{\IfNoValueTF{#3}{Tables~\ref{#1} and \ref{#2}}{\IfNoValueTF{#4}{Tables~\ref{#1}-\ref{#3}}{Tables~\ref{#1}-\ref{#4}}}}}

\DeclareDocumentCommand \fref{ooo} {\IfNoValueTF{#2}{Fig.~\ref{#1}}{\IfNoValueTF{#3}{Figs.~\ref{#1} and \ref{#2}}{Figs.~\ref{#1}-\ref{#3}}}}

\usepackage{graphicx}
\usepackage{multirow}

\usepackage{units, xfrac}

\mathchardef\mhyphen="2D

\usepackage{accents}

\newif\ifShowALL
\newif\ifShowSI

\usepackage[
backend=bibtex,
citestyle=numeric-comp, 
sorting=none, 			
url=false, 				
doi=false, 				
isbn=false, 			
eprint=false, 			
maxcitenames=10, 		
maxbibnames=10,			
firstinits=true
]{biblatex}
\AtBeginBibliography{\small}

\DeclareFieldFormat{labelnumberwidth}{#1\adddot}

\renewbibmacro{in:}{%
	\ifentrytype{article}{}{%
		\printtext{\bibstring{in}\intitlepunct}%
	}%
}	

\renewbibmacro*{volume+number+eid}{
	\printfield{volume}
	\printunit{\addcolon}
}

\renewbibmacro*{journal+issuetitle}{%
	\usebibmacro{journal}%
	\setunit{\adddot\space}%
	\usebibmacro{volume+number+eid}%
}

\DeclareFieldFormat[article]{title}{#1}

\DeclareFieldFormat{date}{#1\setunit{\adddot\space}}

\DeclareFieldFormat{pages}{#1}

\DeclareBibliographyDriver{article}{%
	\usebibmacro{author}%
	\setunit{\adddot\space}\newblock
	\usebibmacro{date}%
	\usebibmacro{title}%
	\usebibmacro{journal}%
	\usebibmacro{volume+number+eid}%
	\usebibmacro{note+pages}%
	\usebibmacro{finentry}%
}

\DeclareFieldFormat[misc]{title}{#1}

\DeclareFieldFormat{url}{\url{#1}\adddot}

\renewbibmacro*{url+urldate}{%
	\ifentrytype{misc}{\usebibmacro{url}}{}%
} 

\DeclareBibliographyDriver{misc}{%
	\usebibmacro{author}%
	\usebibmacro{date}%
	\usebibmacro{title}%
	\usebibmacro{url+urldate}%
	\usebibmacro{finentry}%
}

\begin{document}
	
	\begin{flushleft} 
		{\LARGE \textbf\newline{Quantitatively visualizing bipartite datasets}}
		\newline
		\\
		\textbf{Tal Einav}\textsuperscript{1,}*,
		\textbf{Yuehaw Khoo}\textsuperscript{2},
		\textbf{Amit Singer}\textsuperscript{3}
		\\
		{\small 
			$^{1}\,$Divisions of Computational Biology \& Basic Sciences, Fred Hutchinson Cancer Center, Seattle, WA
			$^{2}\,$Department of Statistics, University of Chicago, Chicago, IL\\
			$^{3}\,$Department of Mathematics and PACM, Princeton University, Princeton, NJ\\
			*$\,$Correspondence: teinav@fredhutch.org
		}
	\end{flushleft}

\section*{Abstract}

As experiments continue to increase in size and scope, a fundamental challenge
of subsequent analyses is to recast the wealth of information into an intuitive
and readily-interpretable form. Often, each measurement only conveys the
relationship between a pair of entries, and it is difficult to integrate these
local interactions across a dataset to form a cohesive global picture. The
classic localization problem tackles this question, transforming local
measurements into a global map that reveals the underlying structure of a
system. Here, we examine the more challenging bipartite localization problem,
where pairwise distances are only available for bipartite data comprising two
classes of entries (such as antibody-virus interactions, drug-cell potency, or
user-rating profiles). We modify previous algorithms to solve bipartite
localization and examine how each method behaves in the presence of noise,
outliers, and partially-observed data. As a proof of concept, we apply these
algorithms to antibody-virus neutralization measurements to create a basis set
of antibody behaviors, formalize how potently inhibiting some viruses
necessitates weakly inhibiting other viruses, and quantify how often
combinations of antibodies exhibit degenerate behavior.

\pagebreak
\section*{Introduction}

Given a country's geographic map, it is straightforward to determine the
distance between any pair of cities. Yet posing this question in reverse (called
the classic localization problem) is far more challenging: given only the
distances between pairs of cities, can we reconstruct the full geographic map
\cite{Kruskal1978}?

Across all scientific disciplines, the interactions between vast numbers of
entries are routinely measured, yet the deeper relationships underlying these
entries only become apparent when recast into a global description of the
system. For geographic maps, large tables of city-city distances are less
interpretable than a 2D map positioning cities relative to one another.

\begin{figure}[t]
	\centering \includegraphics[scale=1]{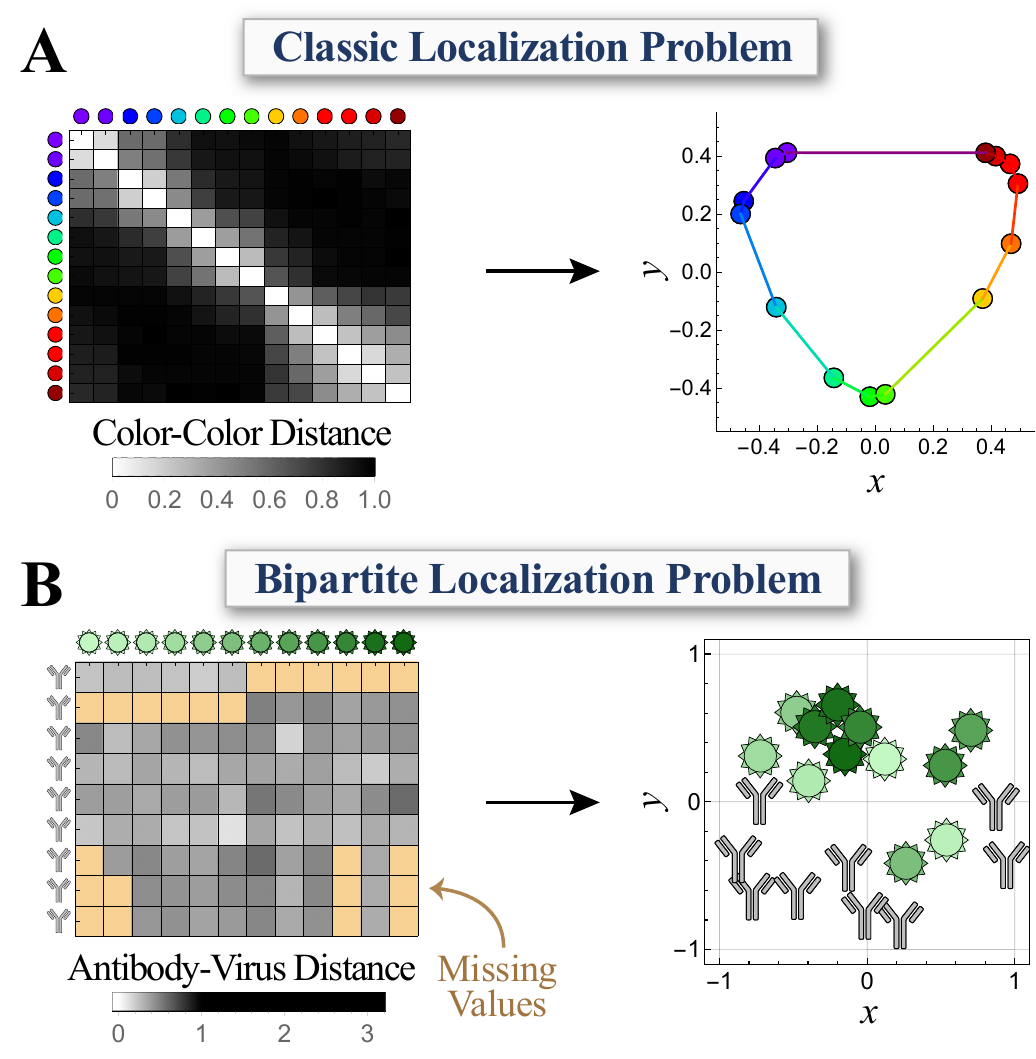}
	
	\caption{\textbf{Embedding monopartite or bipartite data in Euclidean space.}
		(A) The perceived similarity between colors recovers the canonical color wheel.
		Adapted from Table 4.1 of \cite{Borg2005}, with
		distance=1-(dissimilarities in table). (B) Embedding antibody neutralization
		against strains of the influenza virus. In this case, only antibody-virus
		distance can be measured experimentally, and some distances are missing (tan).
		Viruses are colored from lightest-to-darkest hues (oldest to more recent
		strains; full data in Figure~\ref{fig:AnnotatedInfluenzaData}). Adapted from
		Figure 5 of \cite{Creanga2021}, with
		distance=$\log_{10}(\frac{\text{50\% Neutralization}}{10^{-10}
			\text{Molar}})$.} \label{fig:overview MDS}
\end{figure}

To take another example from the field of human perception, the similarity
between pairs of colors reveals that reds, greens, blues, and violets cluster
together (Figure~\ref{fig:overview MDS}A, left). Yet by embedding these
measurements into 2D space (without any additional information about the
colors themselves), the colors naturally form into a highly-intuitive color wheel
(Figure~\ref{fig:overview MDS}A, right). This representation greatly reduces the
complexity of the system, enabling us to hypothesize how new colors would be
perceived and predict trends in the data (\textit{e.g.}, that each color has a
maximally-distant ``complementary color'' on the opposite side of the wheel).

When systems have such a simple underlying structure, we intuitively expect that
a straightforward algorithm can dissect the pairwise distances and recover the
global embedding. Indeed, for complete and noise-free data this can be achieved
in two steps: the first centering the distances to reveal a matrix of inner
products, and a second step using the singular value decomposition to determine
the coordinates (Appendix~\ref{appendix:cMDS}) \cite{Mead1992}. For noisy or
partially-missing data, numeric minimization \cite{DeLeeuw2009, Liberti2014} and
semidefinite programming relaxations \cite{Alfakih1999, Biswas2006,
	Weinberger2006, So2006} have been developed to drive nonlinear dimensionality
reduction \cite{Weinberger2006}, nuclear magnetic resonance spectroscopy
\cite{Wuthrich1990, Donald2011}, and sensor network localization
\cite{Alfakih1999, Biswas2006, So2006, Singer2008, Liberti2014}.

In this work, we consider a twist on this classical problem that we call
\textit{bipartite localization}, where a bipartite dataset consists of two
classes of entries, and interactions can only be measured between (and not
within) each class. Since previous methods are poorly suited to handle bipartite
data \cite{Connelly2017, Gao2022}, we modify existing methods and tailor them
for bipartite localization. In particular, we discuss two variants of the
popular multidimensional scaling (MDS) algorithm -- metric MDS and bipartite MDS
-- as well as a semidefinite programming (SDP) approach \cite{Biswas2006}. Each
method has its own advantages: metric MDS is the simplest and most flexible
numerical framework, bipartite MDS provides a closed-form solution up to an
affine transform, and SDP uses a convex-relaxation that is harder to trap in
local minima. For reproducibility, we create a
\href{https://github.com/TalEinav/Bilocalization}{GitHub repository} with
example code for each algorithm.

Bipartite datasets are ubiquitous in every scientific field, making these
embedding methods broadly applicable. Examples include user-rating profiles such
as the Netflix Challenge \cite{Feuerverger2012, He2016}, graph
clustering \cite{Dhillon2001, Kluger2003, Madeira2004}, the
dimensionality of facial expressions \cite{Russell1985}, the
activity of protein mutants \cite{Jones2019}, gene expression for
different DNA promoters \cite{Yu2020}, and the combinatorics of
ligand signaling \cite{Klumpe2022}.

As a proof of principle, we apply these methods to the pressing issue of
antibody-virus interactions, where multiple antibodies are assessed against
panels of virus mutants (Figure~\ref{fig:overview MDS}B). Unlike many previous
efforts that either exclusively visualized the viruses or the antibodies
\cite{Xie2017, Greaney2021} or required data to be normalized \cite{Smith2004},
we embed both types of entries into a shared space that directly corresponds to
experimental measurements. The resulting map presents a natural context to probe
features such as clustering and to explore the tradeoffs and inherent
constraints of the system. Through these embeddings, we collapse the complexity
of datasets into a readily interpretable and quantitative framework.

\subsection*{The Need for Embedding Algorithms}

Before exploring the algorithms, we motivate the need for such embeddings by
describing several potential applications. To ground this discussion, we suppose
the bipartite classes represent antibodies and viruses (with distances
describing antibody-virus interactions), although these applications generalize
to any bipartite dataset.

First, an embedding combines datasets and predicts unmeasured interactions. For
example, we cannot directly compare an antibody measured against viruses \#1-6
with a second antibody measured against viruses \#7-12 (top two rows in the
Figure~\ref{fig:overview MDS}B dataset). Yet by embedding both antibodies, we
predict their behavior against all viruses in the dataset. Hence, embeddings
represent a form of matrix completion \cite{Candes2009, Hartford2018}.

Second, an embedding defines the intra-class distances between any two viruses
(or two antibodies), a quantity that by definition cannot be directly measured
through antibody-virus interactions. This intra-class distance describes how
differently any antibody can neutralize the two viruses (\textit{i.e.},
essentially quantifying their cross-reactivity). In the limit where two viruses
lie on the same point, they will be neutralized identically by all antibodies;
when the two viruses lie far apart, their neutralization can greatly differ.

Third, the inferred virus-virus distances are crucial when designing future
experiments. Viruses that are close together offer redundant information,
whereas sampling viruses that are spread out across the map can detect more
distinct antibody phenotypes.

Fourth, an embedding defines a basis set of behaviors, which is essential for
systems where no mechanistic models exist. For example, there is a dearth of
models that enumerate the space of antibody behaviors \cite{Smith1999, Wang2015,
	Mayer2015}, which hinders theoretical exploration into features such as the
optimality or degeneracy of the antibody response (both of which we address
later in this work).


Finally, embeddings provide a fundamentally different vantage to study a
system, and this shift in perspective could help uncover its underlying rules.
For example, the complex sequence-to-function relationship of viral proteins may
be simpler to crack within a low-dimensional embedding. Similarly, the antibody
response changes with each viral exposure, and the dynamics of how each antibody
evolves may be more readily understood within the context of an embedding.


%

\section*{Algorithms} 

We next develop the algorithms to transform pairwise interactions into a global
map of a system. In bipartite embedding, we seek to recover the bipartite set of
points $\{x_i^*\}_{i=1}^m, \{y_j^*\}_{j=1}^n\subset \mathbb{R}^d$ given the
noisy distance matrix $D  \in \mathbb{R}^{m\times n}$ of the form
\begin{equation}
	D_{ij} = D_{ij}^* + \epsilon_{ij}, \,\,\, D_{ij}^* = \|x_i^* - y_j^*\|,
\end{equation}
where distance is only measured between the $\{x_i^*\}$ and $\{y_j^*\}$.
$D_{ij}^*$ represents the true distance that is perturbed with independently
and identically distributed random noise $\epsilon_{ij}$. The goal is to use
the noisy $D_{ij}$ with $(i,j)\in \mathcal{E}$, where $\mathcal{E}$ represents the
subset of measured values, to find an embedding $\{x_i\}_{i=1}^m, \{y_j\}_{j=1}^n$ that
approximates the true embedding $\{x_i^*\}, \{y_j^*\}$. In the following
sections, we describe three algorithms to tackle this problem.

\subsection*{Metric Multidimensional Scaling (Metric MDS)}

The straightforward numerical approach is to randomly initialize each $x_i$ and
$y_j$, and then apply numerical methods (\textit{e.g.}, gradient descent or
differential evolution) to match their coordinates as closely as possible to
the distance matrix. In this paper, we use the least-squares loss function
\begin{equation} \label{eq:mMDS loss}
	\min_{\{x_i\}^m_{i=1}, \{y_j\}^n_{j=1}} \sum_{(i,j)\in \mathcal{E}} ( D_{ij} - \norm{x_i-y_j} )^2,
\end{equation}
although we note that other loss functions can strongly affect the embedding
(Figure~\ref{fig:LossFunctionMMDS}). While this method is simple to implement,
it is liable to get trapped in local minima and does not harness the underlying
structure of the bipartite data.
	
\subsection*{Bipartite Multidimensional Scaling (Bipartite MDS)}

In stark contrast, bipartite MDS provides a closed-form solution (up to a rigid
transform) for noise-free and complete data. Although variants of the classical
monopartite problem have been developed to deal with large datasets and noisy
measurements \cite{Peterfreund2021}, to our knowledge this technique has not
been extended to complete bipartite data.

The key insight underlying classic MDS is that the \emph{doubly-centered}
squared-distance matrix is intimately related to the inner products (Gram
matrix) of the embedded points. More precisely, we define the centering matrix that subtracts the mean from any vector,
\begin{equation} \label{eq:CenteringMatrix}
	J_k = I_k - \frac{1}{k} \textbf{1}_k  \textbf{1}_k^T \in \mathbb{R}^{k \times k},
\end{equation}
where $I_k$ is the $k\times k$ identity matrix and $\textbf{1}_k$ is the
all-ones vector of size $k$ (with $J_k \mathbf{1}_k = \mathbf{0}$). Consider the complete
noise-free bipartite graph,
\begin{equation} \label{eq:cMDS D Hadamard}
	(D^*\circ D^*)_{ij} = {D^*_{ij}}^2 = \|x_i^*\|^2 + \|y_j^*\|^2 - 2(x_i^*)^T y_j^*,
\end{equation}
where $\circ$ denotes entrywise multiplication. Double-centering reveals
the inner products of the embedding $X^* = [x_1^*,\ldots,x_m^*]^T \in
\mathbb{R}^{m\times d}$ and $Y^* = [y_1^*,\ldots,y_n^*]^T \in
\mathbb{R}^{n\times d}$ (Appendix~\ref{appendix:BPcMDS}),
\begin{equation} \label{eq:cMDS double centering}
	-\frac{1}{2} J_m (D^*\circ D^*) J_n = J_m X^* (Y^*)^T J_n = X^* (Y^*)^T J_n,
\end{equation}
where in the second equality we assume without loss of generality that the
points in $X^*$ are centered at the origin ($J_m X^* = X^*$).
	
The rank $d$ singular value decomposition (SVD) of the double-centered
squared-distance matrix, $U \Sigma V^T = -\frac{1}{2} J_m (D^*\circ D^*)  J_n$,
determines the embedding of $X^*$ and $Y^*$ up to linear transforms,
\begin{align}
	X^* &= U \Sigma A_U \label{eq:cMDS X} \\
	Y^* &= V A_V + \textbf{1} (t_V)^T, \label{eq:cMDS Y}
\end{align}
for some matrices $A_U, A_V\in \mathbb{R}^{d\times d}$ (satisfying $A_U A_V^T =
I_d$) and a translation $t_V\in \mathbb{R}^d$ between the centers of $X^*$ and
$Y^*$. Lastly, $A_V$ and $t_V$ (together with $A_U = (A_V^T)^{-1}$) are
determined by utilizing the distance information $\norm{ x_i^* - y_j^*} =
D_{ij}^*$ and minimizing~\eqref{eq:mMDS loss} using semidefinite programming or
numeric minimization (Appendix~\ref{appendix:BPcMDS}).

In summary, this algorithm reduces the embedding problem with $(m+n)d$ unknown
variables into the simpler problem of determining the $d^2 + d$ unknown
variables in $A_V$ and $t_V$, \textit{regardless of the size of} $D$! This same
approach can be used for a noisy distance matrix $D$
(Algorithm~\ref{algorithm:cMDS}). A caveat of this method is that it cannot
readily handle missing values. In the numerical experiments below, we first
fill in any missing values using the mean of all observed entries in the same
row and column of the distance matrix -- this leads to poor behavior when a
substantial fraction of values are missing, which can be ameliorated with
metric MDS post-processing (Figure~\ref{fig:ExtramMDS}).
	
\begin{algorithm}[t]
	\setstretch{0.}
	\caption{\textsc{Classical Multidimensional Scaling (Bipartite MDS)}}
	\begin{flushleft}
		\vspace{-0.5em}
		\emph{Input}: 
		\vspace{-0.5em}
		\begin{itemize}
			\vspace{-0.1em}
			\item Distance matrix $D \in \mathbb{R}^{m \times n}$
			\vspace{-0.4em}
			\item Dimension $d$ of the embedding
		\end{itemize}
		\vspace{-0.3em}
		\emph{Steps}:
		\begin{enumerate}
			\vspace{-0.7em}
			\item Define a complete distance matrix $\tilde{D}$ equal to $D$ at measured values, with missing values filled in using the mean of all observed entries in the same row and column
			\vspace{-0.7em}
			\item Compute the double-centered matrix, $Q = -\frac{1}{2} J_m (\tilde{D}\circ\tilde D) J_n$
			\vspace{-0.6em}
			\item Compute the top $d$ SVD, $Q=U \Sigma V^T$
			\vspace{-0.5em}
			\item Set $\{x_i\}_{i=1}^m = U \Sigma A_U$ and $\{y_j\}_{j=1}^n = V A_V + \textbf{1} (t_V)^T$ for linear transforms $A_U,A_V  \in \mathbb{R}^{d \times d}$ and translation vector $t_V  \in \mathbb{R}^{d \times 1}$ (where $A_U A_V^T = I$). Determine $A_U$, $A_V$, and $t_V$ by minimizing the difference between ${D}_{ij}$ and $\norm{x_i-y_j}$ using non-convex numerical minimization or SDP (see Appendix~\ref{appendix:BPcMDS})
		\end{enumerate}
	\end{flushleft} \label{algorithm:cMDS}
	\vspace{-0.5em}
\end{algorithm}

\subsection*{Semidefinite Programming (SDP)}
	
Lastly, we investigate an intermediate algorithm that harnesses the bipartite
nature of the data to perform a more robust numerical search. More precisely,
by forming a positive-semidefinite matrix, we can adapt the sensor network
localization SDP algorithm \cite{Biswas2006} and utilize efficient conic
solvers for bipartite embedding \cite{Toh1998, Sturm1999}. We define the
combined coordinates
$Z = \begin{pmatrix}
	X\\
	Y
\end{pmatrix} 
\in \mathbb{R}^{(m+n) \times d}$ where $X,Y$ store
$\{x_i\}_{i=1}^m,\{y_j\}_{j=1}^n$. We further define the inner product matrix $G \in
\mathbb{R}^{(m+n) \times (m+n)}$ as
\begin{equation}
	Z {Z}^T = \begin{pmatrix}
		X {X}^T & X {Y}^T\\
		Y {X}^T & Y {Y}^T
	\end{pmatrix} \equiv \begin{pmatrix}
		G_{11} & G_{12}\\
		G_{12}^T & G_{22}
	\end{pmatrix} \equiv G,
\end{equation}
so that the squared-distance between $x_i$ and $y_j$ can be entirely written in
terms of the entries of $G$, namely,
\begin{equation}
	\|x_i-y_j\|_2^2= (G_{11})_{ii} - 2 (G_{12})_{ij} + (G_{22})_{jj}.
\end{equation}
Note that we can exactly recast the optimization over $X$ and $Y$ in terms of an
optimization over a positive semidefinite matrix $G$ of rank $d$. The goal is
then to minimize $\sum_{(i,j)\in \mathcal{E}}\vert (G_{11})_{ii} - 2
(G_{12})_{ij} + (G_{22})_{jj} - D_{ij}^2\vert$ in terms of $G$. To this end, we
introduce an extra error matrix $E \in \mathbb{R}^{m \times n}$ and minimize
over the sum of errors:
\begin{equation} \label{eq:SDP Main}
	\begin{matrix*}[l]
		\underset{G,E}{\text{minimize}}		& \sum_{(i,j)\in \mathcal{E}} E_{ij} \\
		\text{subject to}	&E \ge 0, G \succeq 0, \\ 
		& -E_{ij} \le (G_{11})_{ii} - 2 (G_{12})_{ij} + (G_{22})_{jj} - D_{ij}^2 \le E_{ij},\quad (i,j)\in\mathcal{E}\\
		& \sum_{j=1}^n G_{ij} = 0, \,\,\,\, \forall 1 \le j \le m.
	\end{matrix*}
\end{equation}
The final constraint ensures that the $X$ coordinates are centered at the
origin, removing their translational degree of freedom. Note that to achieve
this convex conic program, we removed the non-convex $\text{rank}(d)$ constraint
of $G$, which must now be added back. Thus, we apply an SVD to $G$ of rank $d$,
$G = U \Sigma V^T$. The resulting $m+n$ coordinates are given by
$\begin{pmatrix} X \\ Y \end{pmatrix} = U \sqrt{\Sigma}$ (Algorithm~\ref{algorithm:SDP}).

As with metric MDS, missing values are seamlessly handled in SDP since the objective
in Equation~\ref{eq:SDP Main} is restricted to the measured distances. As shown in the
following sections, SDP often recovers a better embedding than metric or bipartite MDS,
especially when there are many missing values. Note that we specifically chose a
different loss function for metric MDS (Equation~\ref{eq:mMDS loss}, optimized for
systematic noise) and SDP ($\sum_{(i,j)\in \mathcal{E}} \abs{ \norm{x_i-y_j}^2 -
	D_{ij}^2 }$, optimized to handle outliers) in order to explore the diversity of
embedding behaviors. When analyzing datasets, it is worth trying multiple loss
functions to determine which one best characterizes the system
(Figure~\ref{fig:LossFunctionMMDS}C). For completeness, we note that bipartite MDS is
a closed-form method that does not explicitly use any loss function.
	
\begin{algorithm}[t]
	\setstretch{0.}
	\caption{\textsc{Semidefinite Programming (SDP)}}
	\begin{flushleft}
		\vspace{-0.5em}
		\emph{Input}: 
		\vspace{-0.5em}
		\begin{itemize}
			\vspace{-0.1em}
			\item Distance matrix $D \in \mathbb{R}^{m \times n}$
			\vspace{-0.3em}
			\item Dimension $d$ of embedding
		\end{itemize}
		\vspace{-0.3em}
		\emph{Steps}:
		\begin{enumerate}
			\vspace{-0.5em}
			\item Solve $G \in \mathbb{R}^{(m+n) \times (m+n)}$ from Equation~\ref{eq:SDP Main}
			\vspace{-0.5em}
			\item Compute the top $d$ SVD, $G=U \Sigma U^T$. The embedded coordinates $\{x_i\}$ are given by the first $m$ rows of $U \Sigma^{1/2}$ while $\{y_j\}$ are given by the final $n$ rows
		\end{enumerate}
		\vspace{-1.3em}
	\end{flushleft} \label{algorithm:SDP}
\end{algorithm}
	
\section*{Numerical Experiments}

We first assess the three embedding algorithms -- metric MDS , bipartite MDS,
and SDP -- using simulated data with $m=20$ entries $x_i$ and $n=20$ entries
$y_j$ (each chosen uniformly on $[-1,1] \times [-1,1]$). These points generate
the true distance matrix, which we then perturb and use as the input matrix $D$.
The accuracy of the resulting embedding is calculated using the RMSE of
Euclidean distances, $\sqrt{(\sum_{i=1}^m \|x_i - x_i^*\|^2 + \sum_{j=1}^n \|y_j
	- y_j^*\|^2)/(m+n)}$, between the estimated and true coordinates (once aligned
via a rigid transform).
	
\subsection*{Systematic Noise and Missing Values}

\begin{figure}[t]
	\centering \includegraphics[scale=1]{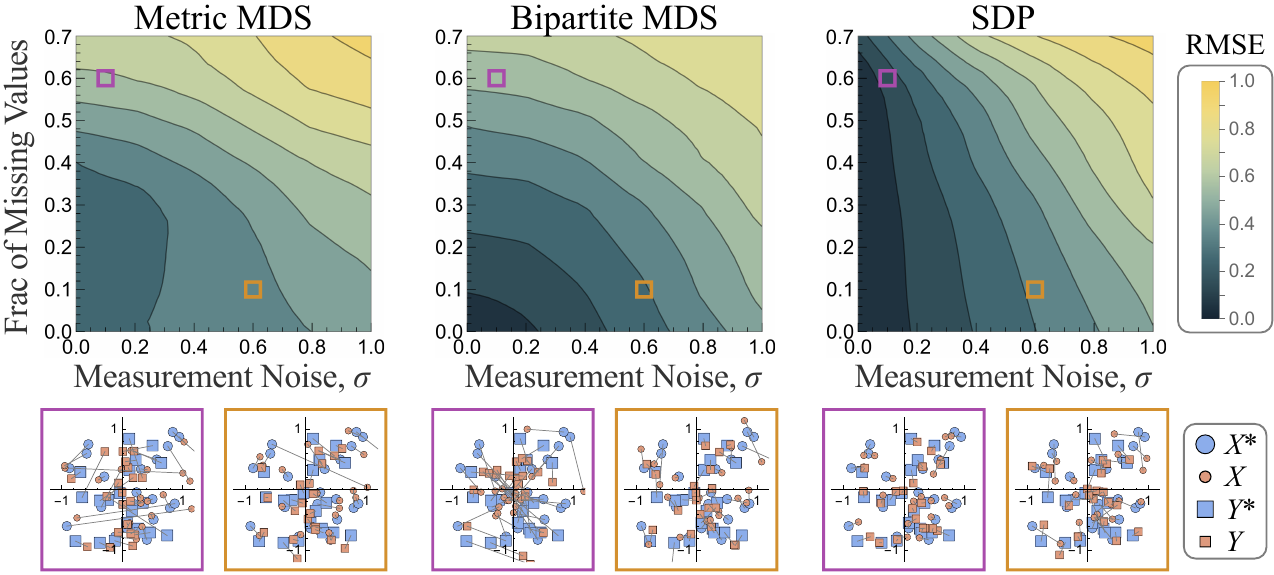}
	
	\caption{\textbf{Performance on a simulated dataset.} \textit{Top}, Phase
		diagram of embedding error as a function of the element-wise noise $\sigma$ of
		the distance matrix and the fraction $f_{\text{Missing}}$ of missing entries
		for metric multidimensional scaling (metric MDS), bipartite multidimensional scaling, and semidefinite programming (SDP). Error is computed as the
		average Euclidean distance between the numerical and true coordinates (aligned
		using a rigid transform). Diagrams show the average of 10 runs, and the metric MDS
		results were smoothed because its embedding accuracy was erratic.
		\textit{Bottom}, Examples of the embedding when $\sigma=0.1$ and
		$f_{\text{Missing}}=0.6$ (purple box) as well as $\sigma=0.6$ and
		$f_{\text{Missing}}=0.1$ (brown box) for each method. Edges connect the
		numerical coordinates to the true embedding.} \label{fig:simulated data}
\end{figure}

To generate the input matrix $D$, we perturb each entry of the true distance
matrix by adding a random value uniformly chosen from $[-\sigma, \sigma]$
($x$-axis) and withhold a fraction $f_{\text{Missing}}$ of randomly selected
entries ($y$-axis). Of the three algorithms, SDP exhibits the most robust
behavior in the presence of missing values (Figure~\ref{fig:simulated data}),
and in the noise-free case along the $y$-axis it undergoes a phase transition
from near-perfect recovery when $f_{\text{Missing}} \le 0.6$ to noisy recovery
(Figure~\ref{fig:SimulatedDataNumEntriesDependence}A). In contrast, the error of
bipartite MDS increases nearly proportionally to $f_{\text{Missing}}$, since
each missing value must be initialized as the row/column mean which effectively
perturbs the distance matrix. Metric MDS also finds poorer embeddings with
larger $f_{\text{Missing}}$, as it occasionally gets trapped in local minima
(even in the low-noise limit).
	
When $D$ is fully observed along the $x$-axis, the error increases approximately
linearly with noise for all three algorithms ($\text{RMSE} \approx \sigma/2$,
Figure~\ref{fig:SimulatedDataNumEntriesDependence}B), although metric MDS
displays somewhat erratic behavior as it may get stuck in local minima. The
bottom panels in Figure~\ref{fig:simulated data} show example embeddings in the
intermediate regimes when $\sigma=0.1$ and $f_{\text{Missing}} = 0.6$ (purple)
or when $\sigma=0.6$ and $f_{\text{Missing}} = 0.1$ (brown), with gray lines
connecting the true coordinates to their numerical approximations.

In terms of overall performance, the region of near-perfect recovery is largest
for SDP followed by bipartite MDS and metric MDS (Figure~\ref{fig:simulated
	data}). One way to improve these algorithms is to combine them, for example, by
using SDP or bipartite MDS to initialize the coordinates in metric MDS. These
combined algorithms substantially improve embedding accuracy, allowing bipartite
MDS to handle missing values and extending the capability of SDP to embed noisy
measurements (Figure~\ref{fig:ExtramMDS}).
	
\subsection*{Handling Large Outliers and Bounded Measurements}

\begin{figure}[t]
	\centering \includegraphics[scale=1]{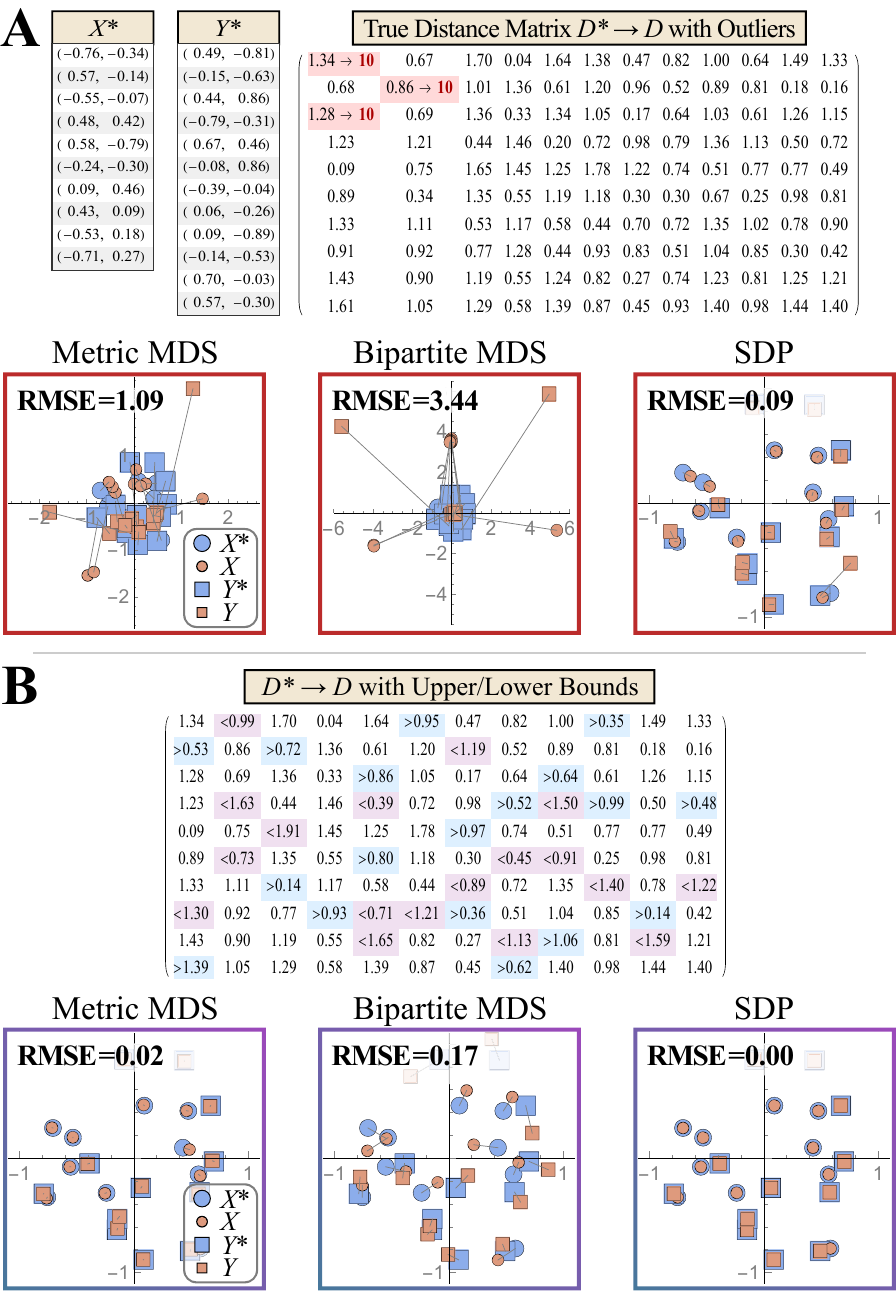}
	
	\caption{\textbf{Embedding with outliers and bounded data.} (A) Embedding a
		noise-free distance matrix $D$ with three highly-corrupted measurements
		(highlighted in red). (B) Embedding a distance matrix where 30\% of entries are
		replaced with upper or lower bounds (blue and purple).} \label{fig:outliers}
\end{figure}
	
In addition to noisy measurements, datasets may contain outliers that distort
an embedding. Bipartite MDS is highly susceptible to large outliers, which can
corrupt the largest singular vectors of the squared-distance matrix
(Figure~\ref{fig:outliers}A). In contrast, SDP minimizes the sum of absolute
(un-squared) deviation \cite{Pollard1991}, and such loss is far more robust
against gross corruptions. Metric MDS exhibits intermediate behavior, although
we note that the choice of loss function heavily influences this behavior
(Figure~\ref{fig:LossFunctionMMDS}).
	
Lastly, we explored each algorithm's tolerance to distances given as
upper or lower bounds, which can arise when an experiment measures a value outside
of its dynamic range. Figure~\ref{fig:outliers}B shows the embedding from the same distance matrix,
now modified to represent 30\% of measurements as upper or lower bounds. In this
complete and noise-free case, both metric MDS and SDP can directly utilize these
bounds to generate near-perfect reconstructions. In contrast, bipartite MDS
cannot directly incorporate bounded data, and hence we replace each bounded
measurement by the bound itself, which leads to worse reconstruction. 



\section*{Analysis of Antibody-Virus Measurements}
	
We next applied these embedding algorithms to an influenza dataset where the
neutralization from 27 stem antibodies was measured against 49 viruses that
circulated between 1933-2019 (Figure~\ref{fig:AnnotatedInfluenzaData}). The
following section transforms these experimental measurements into map distances
to embed these antibody-virus interactions, while all subsequent sections
utilize this embedding to probe the antibody response.
	
\subsection*{Transforming Antibody-Virus Measurements into Distances}

For each antibody-virus pair, the inhibitory concentration required to
neutralize 50\% of virus particles ($\text{IC}_{50}$ in Molar units) was
measured, with lower values signifying a more potent antibody
\cite{Creanga2021}. $\text{IC}_{50}$s ranged from $8.6 \cdot 10^{-11}\,\text{M}$
(very strong neutralization) to $>$$1.6  \cdot 10^{-7}\,\text{M}$ (weak neutralization outside the range of the assay).

To briefly describe the biological context for this dataset, each of the 27
antibodies targets the stem region of hemagglutinin, one of the key surface
proteins on the influenza virus. This stem domain is highly conserved, and
antibodies targeting it can neutralize very diverse viruses; for example, some
antibodies measurably neutralize both the H1N1 and H3N2 influenza subtypes,
which is rarely seen in antibodies targeting the head domain of this same viral
protein \cite{Crowe2018}.

Yet even these broadly neutralizing antibodies have limits. Antibodies that
potently neutralize H1N1 viruses tend to weakly neutralize H3N2 strains (and
vice versa), while antibodies that neutralize all viruses tend to have
intermediate effectiveness. These trends hint that there is an underlying
tradeoff between antibody potency (how much a virus is neutralized) and breadth
(how many diverse viruses can be neutralized). Such patterns are difficult to
directly discern from a table of pairwise interactions, yet they naturally
emerge through an embedding.

\begin{figure}[t]
	\centering \includegraphics[scale=1]{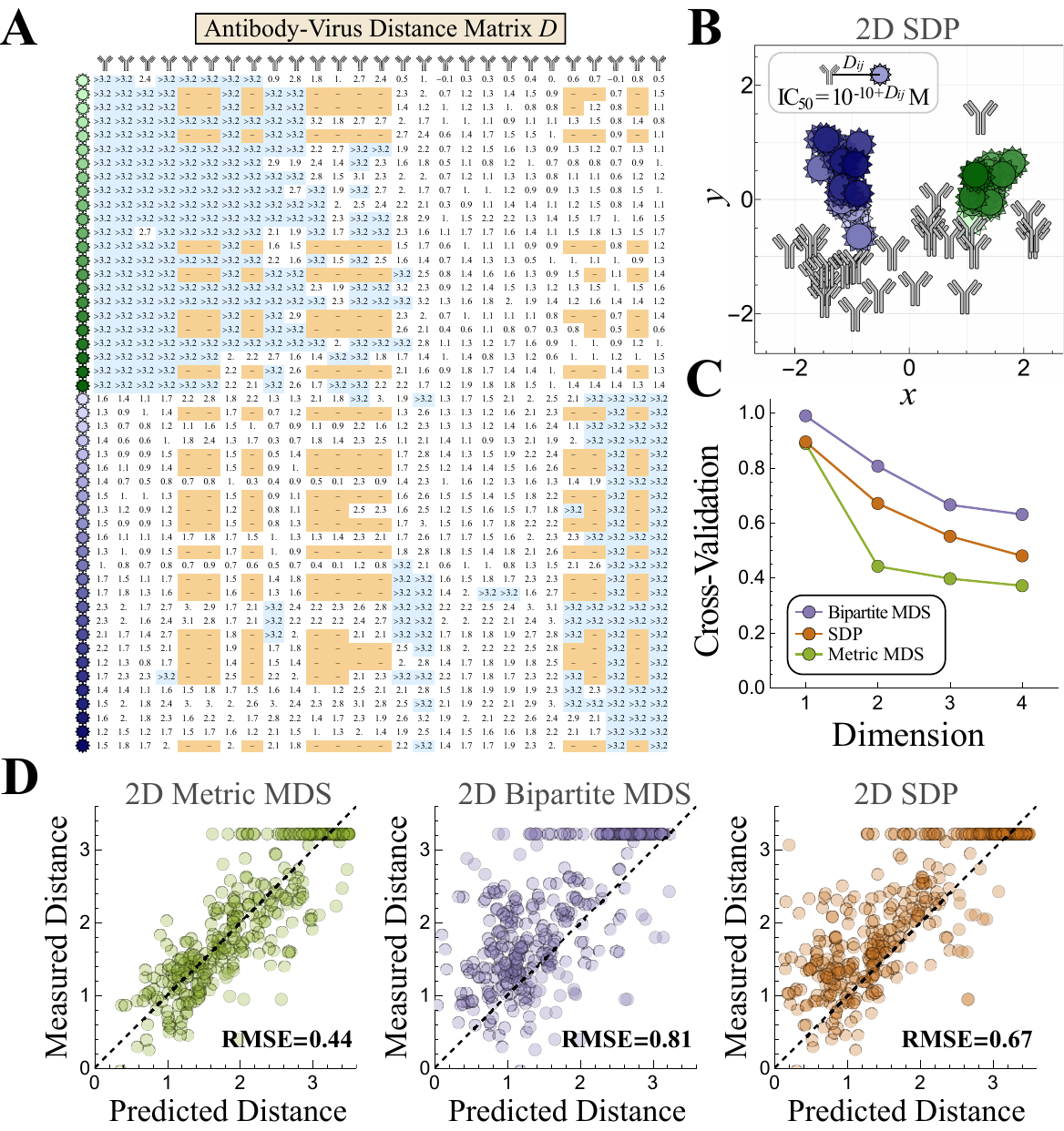}
	
	\caption{\textbf{Mapping influenza antibody-virus interactions.} (A)
	Experimentally measured distance matrix between 27 antibodies and 49 influenza
	viruses \cite{Creanga2021}. (B) The metric MDS embedding in 2D. (C) 10-fold
	cross validation RMSE (calculated using the distance matrix). (D) Example of 2D
	cross validation for each method, demonstrating that metric MDS performs the
	best.} \label{fig:AbVirusData}
\end{figure}

To that end, we first converted these antibody-virus neutralization measurements
into distances. Antibodies typically have
$\text{IC}_{50}$s$\,>10^{-10}\,\text{M}$ (since selection does not act below
this point \cite{Foote1995, Andrews2015}), and hence we define antibody-virus
distance as $D_{ij}=\log_{10} \left( \frac{\text{IC}_{50}}{10^{-10}\,\text{M}}
\right)$ (Figure~\ref{fig:AbVirusData}A). We then applied all three embedding
algorithms to create a global map of the system. Since both the dimensionality
and the ground truth coordinates are not known, we assessed each algorithm
through cross validation (training on 90\% of data, testing on the remaining 10\%).

Metric MDS performed the best in all dimensions and exhibited a sharp ``elbow''
at $d=2$, suggesting that a 2D landscape captures the underlying structure of
the system (Figure~\ref{fig:AbVirusData}C). We note that the 2D cross-validation
RMSE was 0.44 (Figure~\ref{fig:AbVirusData}D), so that withheld neutralization
measurements are predicted within $10^{0.44} \approx 3$-fold, which is
comparable to the noise of the neutralization assay.

\subsection*{Designing Optimal Antibody Cocktails}

The resulting map provides a powerful way to quantify trends in the data
(Figure~\ref{fig:AbVirusData}B). For example, the H1N1 viruses [green] and H3N2
viruses [blue] cluster together, as expected based on their genetic similarity.
Interestingly, the centers of these clusters are $\approx \,$2.5 map units
apart, demonstrating that while antibodies can be highly potent against H1N1 or
H3N2 viruses, no antibody in the panel could strongly neutralize both subtypes.


Similar to the color wheel example in Figure~\ref{fig:overview MDS}A, the
antibody-virus embedding not only represents the entities in this specific
dataset, but also describes other potential antibodies and viruses. For such
entities, the embedding serves as a discovery space to quantify and constrain
their behavior.

For example, within this framework we can design a mixture of $n$ antibodies
that optimally neutralizes the 5 viruses at the top of the H1N1 cluster as well
as the 5 viruses at the top of the H3N2 cluster as potently as possible
(Figure~\ref{fig:MixtureBreathPotency}). This question lies at the heart of
ongoing efforts to find new broadly-neutralizing antibodies, yet few methods
exist to predict or even constrain antibody behavior. To that end, we use each
point on the map to describe a potential antibody whose neutralization against
each mapped virus is determined by its map-distance. This reduces the complex
biological problem of enumerating antibody behavior to a straightforward
geometry problem.

The best $n=1$ antibody mixture against these 10 viruses is
represented by the center of the smallest circle that covers every virus
(Figure~\ref{fig:MixtureBreathPotency}, $\text{distance} \! \le \! 1.4$
[$\text{IC}_{50} \! \le \! 10^{-8.6}\,\text{M}$] for each virus). For a mixture
with $n=2$ antibodies, the potency dramatically improves by using one
H1N1-specific antibody and one H3N2-specific antibody ($\text{distance} \! \le
\! 0.3$ [$\text{IC}_{50} \! \le \! 10^{-9.7}\,\text{M}$] for each virus). This
problem can be readily extended to mixtures with an arbitrary $n$ antibodies
covering any set of mapped viruses. Given the growing number of efforts to find
broadly neutralizing antibodies \cite{Wrammert2011, LeeGeorgiou2016,
	Wagh2018, Andrews2019}, it is essential to have some
framework to estimate the limits of antibody behavior. Such estimations inform
when the antibodies already discovered are near the theoretical best behavior
(and hence further searching is less likely to lead to significant improvement)
or when there are alleged antibodies that could perform orders of magnitude
better than what we have currently seen \cite{Einav2020b}.

\subsection*{Degeneracy of the Antibody Response}

Another key unexplored feature of the antibody response is its degeneracy: can
the neutralization from a mixture of $n$ antibodies behave like a mixture with
fewer antibodies? For example, many vaccination regiments aim to elicit a
broadly-neutralizing antibody that will be potent against diverse viral strains. Yet
even if a post-vaccination antibody response is measured against a large array of
viruses, it may be impossible to determine whether its breadth is conferred by a
single antibody or due to the collective action of multiple antibodies. These
questions hint at an underlying gap in our knowledge, namely, quantifying when
antibody mixtures ``unlock'' fundamentally new behaviors that cannot be achieved
by any individual antibody. Moreover, these topics are difficult to tackle
experimentally, since the low-throughput neutralization assay is time- and
resource-intensive.

Nevertheless, quantifying the degree of antibody degeneracy becomes tractable
through an embedding. Such analyses necessarily make the strong assumption that
every point on the map represents a viable antibody. Moreover, there may be
other antibody phenotypes (\textit{e.g.}, from highly-specific hemagglutinin
head-targeting antibodies) that are not represented by any point on the map; in
essence, the embedding serves to locally extrapolate antibody behavior based on
the specific interactions provided as input (Figure~\ref{fig:AbVirusData}A).
Yet with these caveats, we can explore how often a mixture made within this space of
antibodies can be mimicked by a single antibody.

\begin{figure}[t!]
	\centering \includegraphics[scale=1.]{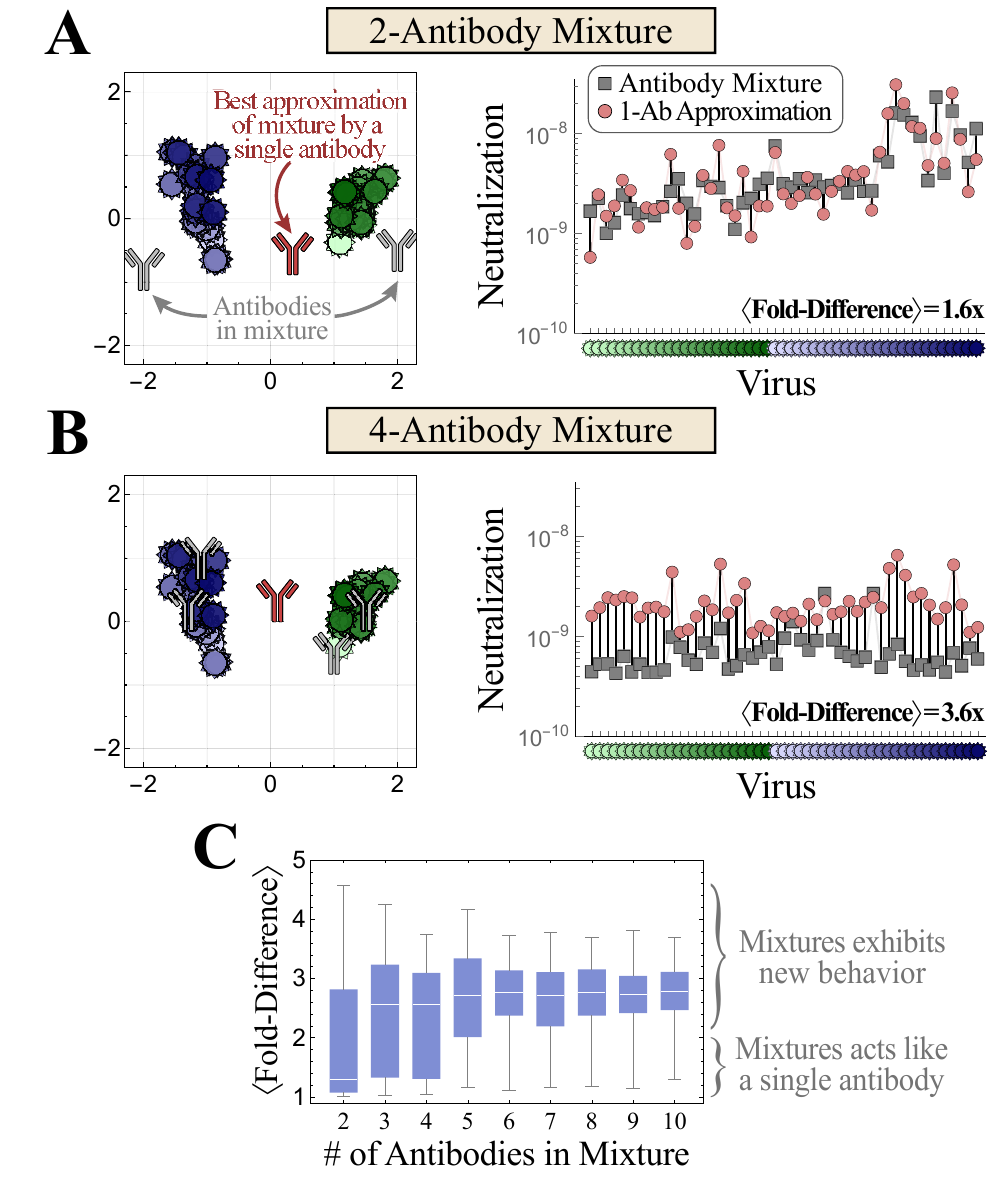}
	
	\caption{\textbf{Degeneracy of antibody mixtures.} Examples of (A) a 2-antibody
		mixture that behaves like a single antibody and (B) a 4-antibody mixture that
		exhibits distinct behavior from any individual antibody. \textit{Left}, the antibodies in
		the mixture (gray) and the best approximating antibody (red). \textit{Right},
		the neutralization $\text{IC}_{50}$s across all viruses. The fold-difference
		between the mixture and antibody is shown by the vertical black lines for each
		virus, with the mean fold-difference given in the bottom-right. (C) For each
		mixture containing $n$ antibodies ($x$-axis), we sample 100 equimolar mixtures
		and quantify their average fold-difference to the nearest approximating
		antibody.} \label{fig:singleAbDecomposition}
\end{figure}

We describe an antibody mixture by $n$ points in Figure~\ref{fig:AbVirusData}B, with the
$i^{\text{th}}$ antibody neutralizing the $j^\text{th}$ virus with an $\text{IC}_{50}^{ij} =
10^{-10 + D_{ij}}$ dictated by the map distance $D_{ij}$ between the antibody
and virus. Since all antibodies in our panel bind to the same region of the
hemagglutinin stem \cite{Joyce2016, Andrews2018, Wu2020}, we treat their binding
as competitive, so only one antibody can bind to each hemagglutinin monomer
at a time. Thus, a mixture's neutralization against virus $j$ is given by
\begin{equation}
	\text{IC}_{50}^{\text{Mixture}} = \left( \sum_i \frac{f_i}{\text{IC}_{50}^{ij}} \right)^{-1}
\end{equation}
where $f_i$ represents the fraction of antibody $i$ in the mixture (with $\sum_i
f_i = 1$). A diluted antibody with small $f_i$ will effectively have a weaker
(larger) $\text{IC}_{50}$, which in the embedding translates to an extra
``distance handicap'' of $\log_{10}f_i$ added to its distance from any virus. We
note that this binding model has been verified on antibody mixtures from this
specific panel \cite{Einav2020b} and on other datasets \cite{Wagh2016,
	Einav2020}. For simplicity, we restrict ourselves to equimolar $n$-antibody
mixtures ($f_i=1/n$).


Given a specific mixture ($n$ random points on the map, sampled near the H1N1
and H3N2 clusters), we quantify the closest approximating single antibody
(another point on the map) by scanning through every possible location and
minimizing the average fold-difference between the mixture's and antibody's
neutralization profiles across all viruses.
Figure~\ref{fig:singleAbDecomposition}A shows a mixture of 2-antibodies (gray),
one of which is potent against the blue H3N2 viruses on the left of the map and
the other potent against the green H1N1 viruses, that behave nearly identically
to a single antibody (red) in the middle of the map. While a few viruses are
neutralized differently by the mixture and antibody (vertical black lines, right
panel of Figure~\ref{fig:singleAbDecomposition}A), on average the antibody's
$\text{IC}_{50}$s are within 1.6-fold of the mixture's values against these 50
diverse viruses. This discrepancy is comparable to the $\approx$2-fold error of
the assay, and hence given either neutralization profile, we could not determine
whether it arose from an individual antibody or a mixture.

Higher-order mixtures unlock more unique behaviors that cannot be replicated by
an individual antibody. For example, not only does the 4-antibody mixture in
Figure~\ref{fig:singleAbDecomposition}B show a 3.6-fold difference from the
nearest approximating antibody, but the mixture's measurements are
systematically lower across nearly all viruses. Thus, neutralization profiles
exhibiting such strong breath are indicative of multiple antibodies.

To systematically explore degeneracy, we sampled 100 antibody mixtures for each
$n$ (with $2 \le n \le 10$) and found the closest approximating single antibody.
The resulting distributions of the mean fold-difference are shown in
Figure~\ref{fig:singleAbDecomposition}C. While 2-antibody mixtures tend to
resemble individual antibodies, higher order mixtures often exhibit distinctive
profiles with a $\langle \text{fold-difference} \rangle > 2$ to the closest
approximating antibody. By the time $n \ge 5$ antibodies are combined, the
likelihood that they match any single antibody becomes exceedingly rare. 

\section*{Discussion}



Embedding algorithms fill a ``hole'' in our understanding by transforming local
pairwise interactions into a global map. Such algorithms have been used to
identify when a new viral variant arises, quantify drug-protein interactions,
and distinguish between cell types \cite{Smith2004, Zitnik2018, Cortal2021}. Yet
we propose that such algorithms also provide the groundwork for new theoretical
studies that only become possible when we reveal the underlying structure of a
system.

In the context of antibody-virus interactions, an embedding provides a rigorous
approach to extrapolate available measurements. Each point describes a potential
antibody, and the entire map defines a basis set of antibody behaviors. By
coupling these data-driven results with a biophysical model of how antibodies
collectively act, we can model higher-order mixtures and pave the way to study
the complex array of antibodies within each person. 

More work is needed to understand the limits of these embeddings and quantify
their predictive power. At the same time, we are just beginning to scratch
the surface on aspects of the antibody response that can be probed with these
embeddings, from designing antibody cocktails to determining how the antibody
response evolves on the map with each viral exposure.

As datasets continue to grow in size and complexity, it becomes increasingly
important to quantitatively visualize interactions between entities. Future
datasets may require multi-localization, where higher-order interactions
(\textit{e.g.}, between a ligand and multimeric receptor \cite{Klumpe2022};
antibodies, antigens, and cell receptors \cite{Tan2021}; or single-cell
multi-omics datasets \cite{Chen2022}) are embedded in a low-dimensional space.

\section*{Acknowledgments} 

We thank Yuval Kluger for his input on this manuscript. Tal Einav is a Damon
Runyon Fellow supported by the Damon Runyon Cancer Research Foundation (DRQ
01-20). Yuehaw Khoo is supported by NSF DMS-2111563. Amit Singer is supported in
part by AFOSR FA9550-20-1-0266, the Simons Foundation Math+X Investigator Award,
NSF BIGDATA Award IIS-1837992, NSF DMS-2009753, and NIH/NIGMS 1R01GM136780-01.

\printbibliography

\pagebreak
\appendix

\renewcommand{\thepage}{S\arabic{page}}
\renewcommand{\thefigure}{S\arabic{figure}}
\renewcommand{\thetable}{S\arabic{table}}
\renewcommand{\theequation}{S\arabic{equation}}
\setcounter{page}{1}
\setcounter{figure}{0}
\setcounter{table}{0}
\setcounter{equation}{0}

\newpage

\section{Supplemental Methods}

The sections below describe the implementations of metric MDS, bipartite MDS, and SDP
algorithms. Complete Mathematica code is available on GitHub
(\href{https://github.com/TalEinav/Bilocalization}{https://github.com/TalEinav/Bilocalization}).

\subsection{Solving the classical localization problem for complete, noise-free data} \label{appendix:cMDS}

Here, we present the well-known solution to the (monopartite) classical
localization problem, where the noise-free distances $D  \in \mathbb{R}^{n
	\times n}$ is provided between every pair of $n$ points in $d$ dimensions. Our
goal is to determine the coordinates $\{x_i\}_{i=1}^n \subset \mathbb{R}^d$ such
that $D_{ij} = \norm{x_i-x_j}$. 

Since an embedding always has a translational degree of freedom, we will assume
without loss of generality that the coordinates are centered around the origin,
$\sum_{i=1}^n x_i = \mathbf{0}$. We define the combined coordinate matrix $X =
[x_1,\ldots,x_n]^T \in \mathbb{R}^{n \times d}$. Note that the entrywise
squared-distance matrix can be written as
\begin{equation}
	D \circ D = \text{diag}(X X^T) \mathbf{1}_n^T + \mathbf{1}_n \text{diag}(X X^T)^T - 2 X X^T
\end{equation}
where the first two terms on the right-hand side are outer products.

The algorithm proceeds in two steps. First, we apply the centering matrix $J_n$ (Equation~\ref{eq:CenteringMatrix}) from the left and right to
row-center and column-center the squared-distances,
\begin{equation}
	-\frac{1}{2} J_n (D \circ D)  J_n = X X^T,
\end{equation}
where we used the fact that $J_n \mathbf{1}_n = \mathbf{0}$ and $\mathbf{1}_n^T
J_n = \mathbf{0}$. This transforms the distance matrix into a matrix of inner
products for the coordinates, $X X^T$.

The second step is to compute the SVD of the left-hand side, $U \Sigma U^T = X
X^T$, which will only have $d$ non-zero singular values. From this form, we can
immediately read out the solution $X = U \Sigma^{1/2}$.

\subsection{Handling missing values in the distance matrix}

In metric MDS and SDP, missing values are automatically ignored, since the sums in
\ref{eq:mMDS loss} and \ref{eq:SDP Main} are over the measured distances,
$(i,j)\in \mathcal{E}$. In other words, each measured edge constrains the
embedding while unmeasured edges are ignored. Bipartite MDS requires a complete matrix to
compute the SVD, and hence each missing entry is first filled in using the mean
of all non-missing measurements in its row and column.

\subsection{Handling upper or lower bounds in the distance matrix}

Sometimes measurements are given as upper or lower bounds on $\|x_i-y_j\|$
(signifying weak or strong interactions outside the dynamic range of the
experiment). 

For an upper bound $\norm{x_i-y_j} < b_\text{up}$ in metric MDS, we modify the
relevant summand in the loss function to $\frac{1}{1+e^{c ( b_\text{high} -
		\norm{x_i-y_j} )}} \left( b_\text{high} - \norm{x_i-y_j} \right)^2$ where $c>0$
is a positive constant (in this work, we chose $c=10$ based on the scale of the
distance measurements). The prefactor in the summand penalizes violations of the
bound while minimally increasing the loss when the bound is satisfied. For a
lower bound, $\norm{x_i-y_j} > b_\text{low}$, we similarly modify the summand to
$\frac{1}{1+e^{-c ( b_\text{low} - \norm{x_i-y_j} )}} \left( b_\text{low} -
\norm{x_i-y_j} \right)^2$. Note that the resulting cost function is non-convex,
which can prevent numerical algorithms from finding a good minimizer. When
computing RMSE in the cross-validation analysis
(Figure~\ref{fig:AbVirusData}C,D), we added these same prefactors when the
measured distance was an upper or lower bound.

Bipartite MDS requires exact distance measurements to compute an SVD, and
hence we replace each bounded measurement by the bound itself, which leads to
poorer embeddings.

In SDP, bounded values are handled by modifying the second constraint in
Problem~\ref{eq:SDP Main}. For example, an upper bound $\|x_i-y_j\| <
b_\text{up}$ is enforced by the one-sided constraint $(G_{11})_{ii} - 2
(G_{12})_{ij} + (G_{22})_{jj} - b_\text{up}^2 < E_{ij}$. A lower bound
$\|x_i-y_j\| > b_\text{low}$ is enforced by the one-sided constraint $-E_{ij} <
(G_{11})_{ii} - 2 (G_{12})_{ij} + (G_{22})_{jj} - b_\text{low}^2$. The objective
remains the same, namely, to minimize the sum of (positive) errors $E_{ij}$
between the embedding and distance measurements.

\subsection{Determining the affine transformation in Bipartite MDS} \label{appendix:BPcMDS}

In this section, we provide some intuition for bipartite MDS and describe in detail the final SDP step that determines the affine transform between $X$ and $Y$.

We begin by rewriting Equation~\eqref{eq:cMDS D Hadamard} as
\begin{equation}
	{D^*} \circ D^* = \text{diag}(X^* {X^*}^T) \mathbf{1}_n^T + \mathbf{1}_m \text{diag}(Y^* {Y^*}^T)^T - 2 {X^*}^T Y^*.
\end{equation}
Because $\mathbf{1}_m,\mathbf{1}_n$ lie
in the nullspace of $J_m, J_n$, double-centering isolates the inner product term
as in Equation~\eqref{eq:cMDS double centering}. Using the rank-$d$ SVD $U
\Sigma V^T = -\frac{1}{2} J_m (D^*\circ D^*)  J_n$, we can rewrite
Equation~\eqref{eq:cMDS double centering} as
\begin{equation}
	X^* = U \Sigma (V^T ({Y^*}^T J_n)^\dagger),\quad J_n Y^* = V (\Sigma U^T ({X^*}^T)^\dagger).
\end{equation}
where ``$\dagger$'' denotes the pseudo-inverse. This reveals that the embedding $X^*, Y^*$ can be determined up to linear transforms as in Equation~\eqref{eq:cMDS X}-\eqref{eq:cMDS Y}.

\begin{figure}[t]
	\centering \includegraphics[scale=1]{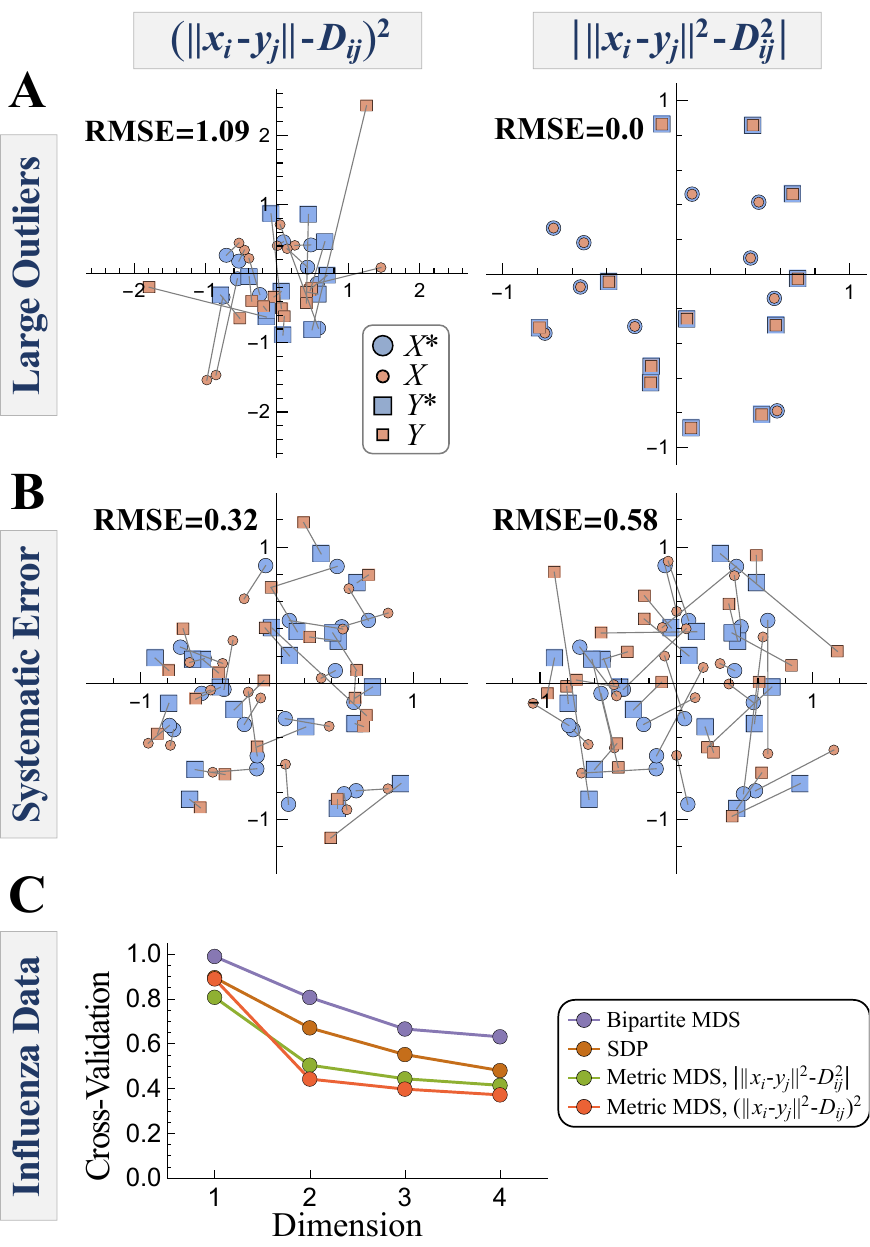}
	
	\caption{\textbf{Choice of loss function strongly influences metric MDS.} We compare
		two loss functions for metric MDS: \textit{Left column} mean squared error between
		unsquared-distances (shown in the main text) versus \textit{right column} mean
		absolute error between the squared-distances. (A) Mean absolute error handles
		the distance matrix with large outliers far better (see
		Figure~\ref{fig:outliers}A). (B) Large systematic noise is handled better by
		mean squared error, since this represents the maximum likelihood estimator for
		approximately-Gaussian error (see $\sigma=1,f_{\text{Missing}}=0$ from
		Figure~\ref{fig:simulated data}). (C) Cross-validation for the influenza
		data in Figure~\ref{fig:AbVirusData} is slightly lower for metric MDS with mean
		squared error for embeddings with $\text{dimension} \ge 2$.}
	\label{fig:LossFunctionMMDS}
\end{figure}

As described in the main text, the final step of bipartite MDS is to determine the
affine transforms $A_U, A_V$ (satisfying $A_U A_V^T = I_d$) and the translation
$t_V$ so that the embeddings $X^*,Y^*$ in Equation~\eqref{eq:cMDS
	X}-\eqref{eq:cMDS Y} match the distance matrix. This can be done numerically by
either minimizing the loss function in Equation~\eqref{eq:mMDS loss} that
optimally handles systematic noise, or by minimizing the loss function of
squared-distances
\begin{equation} \label{eq:SDP loss}
	\min_{\{x_i\}^m_{i=1}, \{y_j\}^n_{j=1}} \sum_{(i,j)\in \mathcal{E}} \abs{ {D}_{ij}^2 - \norm{x_i-y_j}^2 }
\end{equation}
that better handles outliers (Figure~\ref{fig:LossFunctionMMDS}).

Instead of numeric minimization, we can use semidefinite programming to solve
Equation~\eqref{eq:SDP loss} and prevent the minimization from getting stuck at
local minimum (note that with semidefinite programming cannot be used with the
loss function in Equation~\eqref{eq:mMDS loss}). To that end, we construct the
$(2d+1) \times (2d+1)$ Gram matrix
\begin{equation}
	\tilde{G} = \begin{pmatrix}
		- \text{ } A_U \text{ } - \\
		- \text{ } A_V \text{ } - \\
		- \,\,\, t_V^T \,\,\, -
	\end{pmatrix} \begin{pmatrix}
		| & | & | \\
		A_U^T & A_V^T & t_V \\
		| & | & |
	\end{pmatrix}
\end{equation}
with which we can express $\norm{x_i-y_j}^2$. Using Equations~\eqref{eq:cMDS
	X}-\eqref{eq:cMDS Y}, we can write
\begin{equation}
	\norm{x_i-y_j}^2 = U_i \Sigma A_U A_U^T \Sigma U_i^T + V_j A_V A_V^T V_j^T + t_V^T t_V - 2 V_j \Sigma U_i - 2 U_i \Sigma A_U t_V + 2 V_j A_V t_V
\end{equation}
in terms of the entries of $\tilde{G}$.

Define the minimization matrix $\gamma \in \mathbb{R}^{m \times n}$ with
$\gamma_{ij} = {D}_{ij}^2 - \norm{x_i-y_j}^2$. To minimize the absolute value of
the $\gamma_{ij}$, we use Schur's complement condition, defining the auxiliary
matrix $\tilde{\gamma} \in \mathbb{R}^{m \times n}$ and using semidefinite
programming to solve
\begin{equation} \label{eq:cMDS final SDP step}
	\begin{matrix*}[l]
		\underset{\tilde{G}, \tilde{\gamma}}{\text{minimize}}		& \sum_{(i,j)\in \mathcal{E}} \tilde{\gamma}_{ij} \\
		\text{subject to}	&\tilde{G} \succeq 0, \\
		& \begin{pmatrix}
			\tilde{\gamma}_{ij}	&	\gamma_{ij} \\
			\gamma_{ij}			&	1 
		\end{pmatrix} \succeq 0, \,\,\,\, \forall 1 \le i \le m, 1 \le j \le n \\
		& \tilde{G}_{d+1:2d,1:d} = I_d.
	\end{matrix*}
\end{equation}
We then use Cholesky decomposition extract $A_U$ from $\tilde{G}_{1:d,1:d}$ and
determine $A_V = \tilde{G}_{d+1:2d,d+1:2d} A_U$ as well as $t_V =
\tilde{G}_{2d+1, d+1:2d} A_U$. The resulting embedding is given by
Equations~\eqref{eq:cMDS X}-\eqref{eq:cMDS Y}.

\subsection{Numeric Minimization in Metric Multidimensional Scaling}

Minimization was performed using the default NMinimize in \textit{Mathematica}
with the default search method (Differential Evolution). This method is
time-constrained, so that for large distance matrices (where $n+m \gtrsim 100$),
it may yield very poor solutions. Although in this work we only ran this
minimization once, multiple initial conditions could be run to report the lowest
error achieved.

\begin{figure}[t]
	\centering \includegraphics[scale=1]{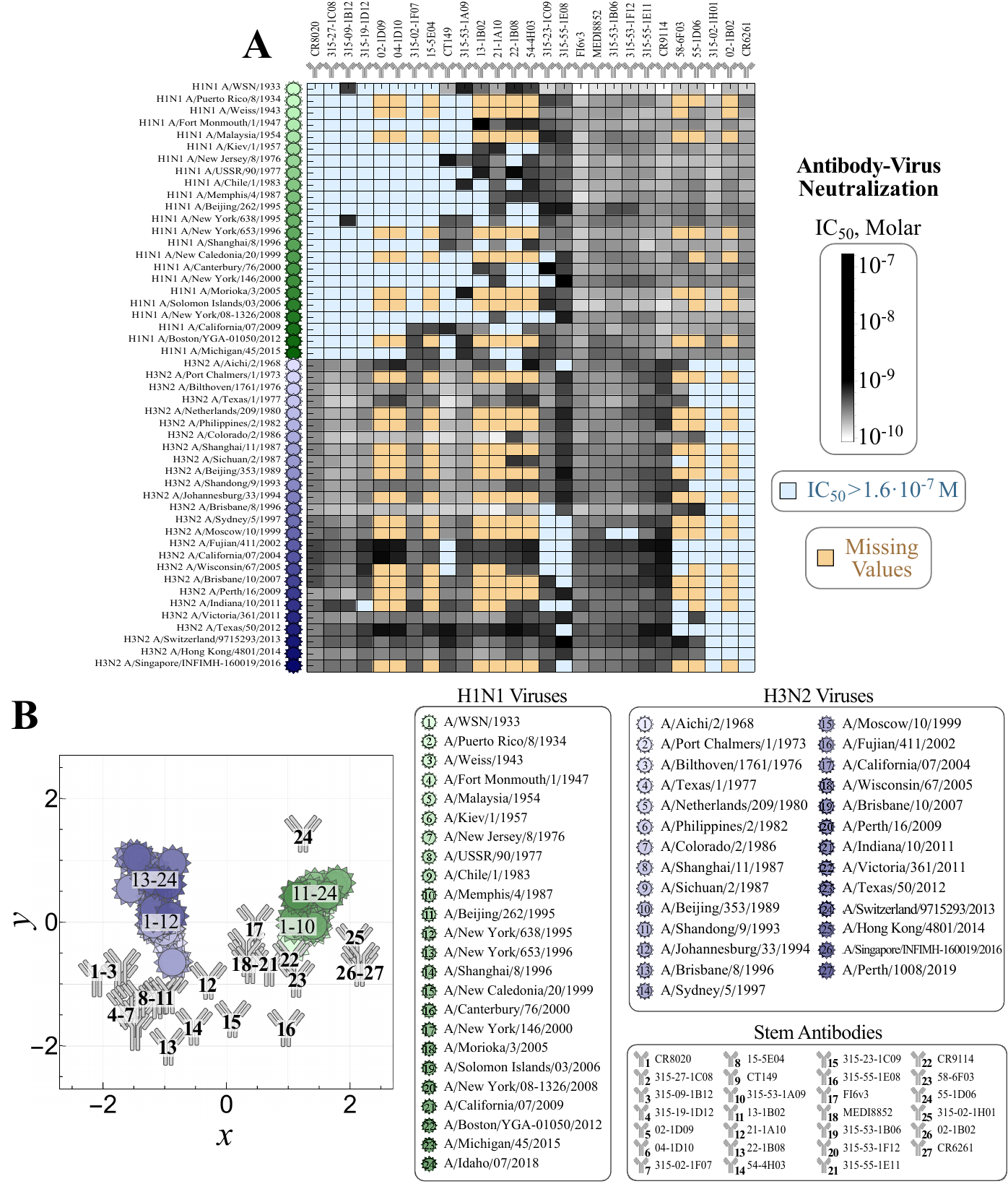}
	
	\caption{\textbf{Annotated influenza antibody-virus data from Creanga
			\textit{et al}. \cite{Creanga2021}.} (A) Neutralization
		measurements of 49 influenza viruses against 27 antibodies targeting the stem
		of influenza hemagglutinin. The inhibitory concentration of antibody needed to
		neutralize 50\% of viruses ($\text{IC}_{50}$, grayscale). Some antibody-virus
		interactions were not measured (tan), and some antibodies exhibited weak
		neutralization ($\text{IC}_{50} > 1.6 \cdot 10^{-7}\,\text{M}$, light-blue)
		outside the dynamic range of the assay. (B) The same 2D metric MDS embedding (as in
		Figure~\ref{fig:AbVirusData}B) with the antibodies and viruses labeled.}
	\label{fig:AnnotatedInfluenzaData}
\end{figure}

\begin{figure}[t]
	\centering \includegraphics[scale=1]{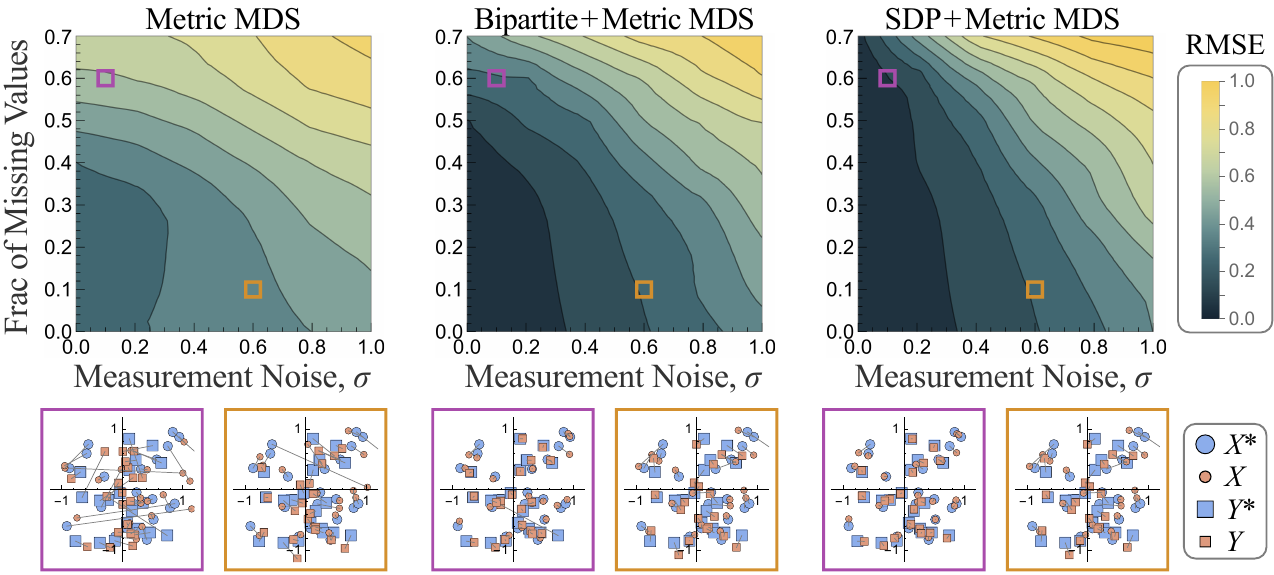}
	
	\caption{\textbf{Post-processing an embedding with metric MDS.} As in
		Figure~\ref{fig:simulated data}, data is simulated with element-wise noise
		$\sigma$ and a fraction $f_{\text{Missing}}$ of missing entries. The results of
		each embedding is used to initialize one additional metric MDS, which greatly
		improves its accuracy. Error is computed as the average Euclidean distance
		between the numerical and actual coordinates (aligned using a rigid transform).
		Example plots at the bottom show an embedding when $\sigma=0.1$ and
		$f_{\text{Missing}}=0.6$ (purple box) as well as $\sigma=0.6$ and
		$f_{\text{Missing}}=0.1$ (brown box) for each method.} \label{fig:ExtramMDS}
\end{figure}

\begin{figure}[t]
	\centering \includegraphics[scale=1]{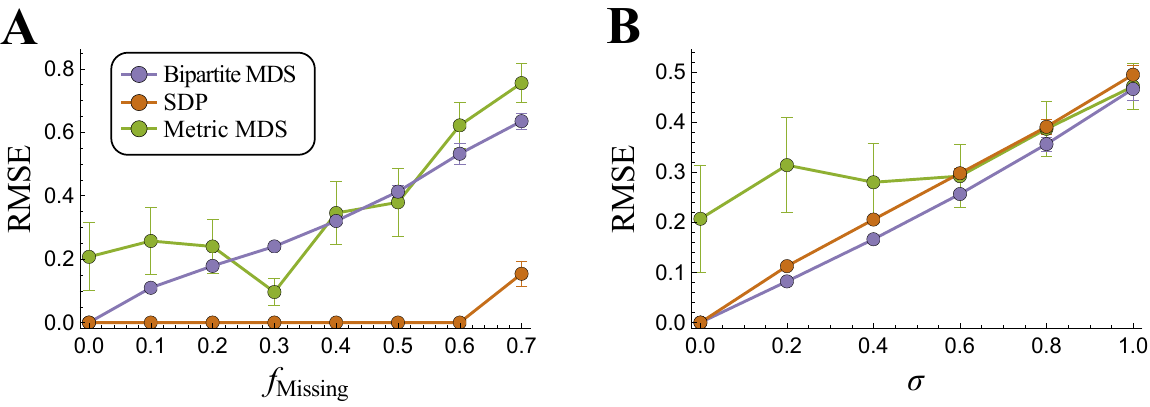}
	
	\caption{\textbf{Embedding for noise-free or complete datasets.} (A) The
		noise-free limit $\sigma=0$ and (B) the limit of complete data
		$f_{\text{Missing}}=0$. In both panels, $m=n=20$ and the error represents the
		RMSE of Euclidean distances between the the estimated and true coordinates
		(once aligned via a rigid transform). }
	\label{fig:SimulatedDataNumEntriesDependence}
\end{figure}

\begin{figure}[t]
	\centering \includegraphics[scale=1]{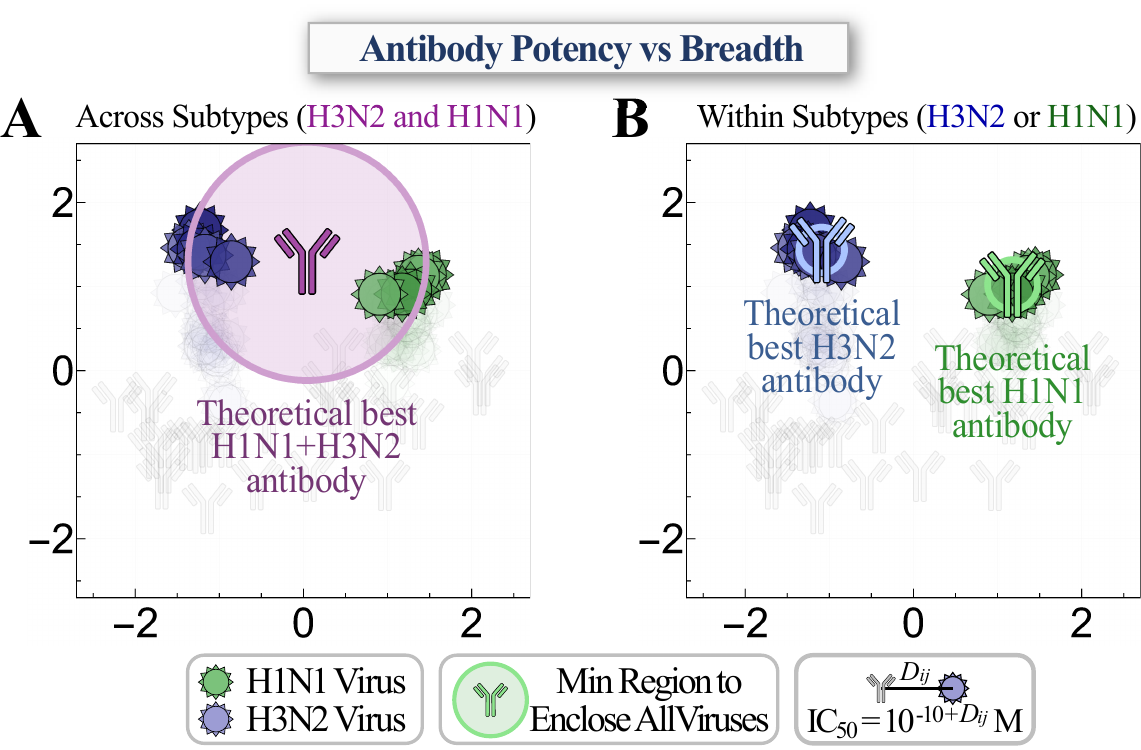}
	
	\caption{\textbf{Predicting optimal neutralization for any mapped viruses.}
		Suppose we want to neutralize the five viruses at the tops of the H1N1 and H3N2
		clusters in Figure~\ref{fig:AbVirusData}B (H1N1 A/New York/638/1995,
		A/Beijing/262/1995, H1N1 A/New Caledonia/20/1999, A/Canterbury/76/2000, A/New
		York/146/2000 and H3N2 A/Fujian/411/2002, A/California/07/2004,
		A/Indiana/10/2011, A/Texas/50/2012, A/Perth/1008/2019). Using the points on the
		map to represent potential antibody neutralization profiles, we determine (A)
		the best single antibody or (B) the best 2-antibody mixture that would
		neutralize these viruses the most potently (with the smallest possible distance
		between any virus and the nearest antibody). The solution is given by the $n$
		minimum covering circles for these viruses, with the antibodies positioned at
		the centers of each circle.} \label{fig:MixtureBreathPotency}
\end{figure}	
	
\end{document}